\documentstyle[12pt]{article}
\newcommand\be{\begin{equation}}
\newcommand\ee{\end{equation}}
\newcommand\bea{\begin{eqnarray}}
\newcommand\eea{\end{eqnarray}}
\newcommand\ket[1]{|#1\rangle}

\newcommand{\fatalpha}{{\bf \alpha \kern -0.44em \alpha}}
\newcommand{\fatsigma}{{\bf \sigma \kern -0.54em \sigma}}
\newcommand{\tpchi}{{\bf \chi \kern -0.35em \chi}}
\newcommand{\llambda}{{\bf \lambda \kern -0.45em \lambda}}

\renewcommand{\theequation}{\arabic{equation}}
\renewcommand{\theequation}{\thesection-\arabic{equation}}
\bibliography{plain}
\title{\bf Perfect state transfer of a qudit over underlying networks of group association schemes}\vspace{20mm}
\author{ M. A. Jafarizadeh$^{a,b,c}$
 \thanks{E-mail:jafarizadeh@tabrizu.ac.ir},
 R. Sufiani$^{a,b}$
 \thanks{E-mail:sofiani@tabrizu.ac.ir}, S. F. Taghavi$^{a}$ and E. Barati$^{a}$
 \\ $^a${\small Department of Theoretical Physics and Astrophysics,
University of Tabriz, Tabriz 51664, Iran.} \\ $^b${\small
Institute for Studies in Theoretical Physics and Mathematics,
Tehran 19395-1795, Iran.} \\ $^c${\small Research Institute for
Fundamental Sciences, Tabriz 51664, Iran.}} \pagebreak

\vspace{20mm}
\begin{document}
\maketitle \vspace{15mm}
\newpage
\begin{abstract}
As generalizations of results of Christandl et al.\cite{8,9''} and
Facer et al.\cite{Facer}, Bernasconi et al.\cite{godsil,godsil1}
studied perfect state transfer (PST) between two particles in
quantum networks modeled by a large class of cubelike graphs
(e.g., the hypercube) which are the Cayley graphs of the
elementary abelian group $Z_2^n$. In Refs. \cite{PST,psd},
respectively, PST of a qubit over distance regular spin networks
and optimal state transfer (ST) of a $d$-level quantum state
(qudit) over pseudo distance regular networks were discussed,
where the networks considered there were not in general related
with a certain finite group. In this paper, PST of a qudit over
antipodes of more general networks called underlying networks of
association schemes, is investigated. In particular, we consider
the underlying networks of group association schemes in order to
employ the group properties (such as irreducible characters) and
use the algebraic structure of these networks (such as Bose-Mesner
algebra) in order to give an explicit analytical formula for
coupling constants in the Hamiltonians so that the state of a
particular qudit initially encoded on one site will perfectly
evolve to the opposite site without any dynamical control. It is
shown that the only necessary condition in order to PST over these
networks be achieved is that the centers of the corresponding
groups be non-trivial. Therefore, PST over the underlying networks
of the group association schemes over all the groups with
non-trivial centers such as the abelian groups,
the dihedral group $D_{2n}$ with even $n$, the Clifford group $CL(n)$ and all of the $p$-groups can be achieved. \\


{\bf PACs Index: 01.55.+b, 02.10.Yn }
\end{abstract}

\vspace{70mm}
\newpage
\section{Introduction}
The quantum communication between two parts of a physical unit,
e.g., a qubit, is a crucial ingredient for many quantum
information processing protocols \cite{1}. There are various
physical systems that can serve as quantum channels, one of them
being a quantum spin system. In view of applications like the
communication between registers in quantum devices, the study of
natural evolution of permanently coupled spin networks has become
increasingly important. A special case of interest consists of
homogenous networks of particles coupled by constant and fixed
(nearest-neighbor) interactions. An important feature of these
networks is the possibility of faithfully transferring a qubit
between specific particles without tuning the couplings or
altering the network topology. This phenomenon is usually called
perfect state transfer (PST). Quantum communication over short
distances through a spin chain, in which adjacent qubits are
coupled by equal strength has been studied in detail, and an
expression for the fidelity of quantum state transfer has been
obtained \cite{Bose,5}. Similarly, in Ref. \cite{6}, near perfect
state transfer was achieved for uniform couplings provided a
spatially varying magnetic field was introduced. After the work of
Bose \cite{Bose}, in which the potentialities of the so-called
spin chains have been shown, several strategies were proposed to
increase the transmission fidelity \cite{Os} and even to achieve,
under appropriate conditions, perfect state transfer
\cite{8,9'',Bu,Bu1,yung,yung1, Facer}. Recently, A. Bernasconi, et
al. \cite{godsil} have studied PST between two particles in
quantum networks modeled by a large class of cubelike graphs.
Since quantum networks (and communication networks in general) are
naturally associated with undirected graphs, there is a growing
amount of literature on the relation between graph-theoretic
properties and properties that allow PST \cite{godsil1}. In Ref.
\cite{PST}, the so called distance-regular graphs have been
considered as spin networks (in the sense that with each vertex of
a distance-regular graph a qubit or a spin $1/2$ particle was
associated) and PST over them has been investigated. In Ref.
\cite{karim}, state transfer over spin chains of arbitrary spin
has been discussed so that an arbitrary unknown qudit be
transferred through a chain with rather a good fidelity by the
natural dynamics of the chain. In the recent paper \cite{psd}, the
authors have investigated optimal state transfer (ST) of a
$d$-level quantum state (qudit) over pseudo distance regular
networks, where it was shown that only for pseudo distance regular
networks with some certain symmetry (mirror symmetry) in the
corresponding intersection numbers (consequently their QD
parameters), PST between antipodes of the networks can be
achieved.

In the present paper we will consider the more general graphs that
are the underlying graphs of group association schemes
\cite{Ass.sch.} and give necessary and sufficient conditions for
PST of a qudit state in quantum networks modeled by a large class
of graphs with group structure. An association scheme is a
combinatorial object with useful algebraic properties (see
\cite{godsil2} for an accessible introduction). The theory of
association schemes has its origin in the design of statistical
experiments. This mathematical object has very useful algebraic
properties which enables one to employ them in algorithmic
applications such as the shifted quadratic character problem
\cite{Childs1} (in this problem, the association scheme is the
Paley scheme which corresponds to a strongly regular graph, the
Paley graph). A $d$-class symmetric association scheme ($d$ is
called the diameter of the scheme) has $d+1$ symmetric relations
$R_i$ which satisfy some particular conditions. Each non-diagonal
relation $R_i$ can be thought of as the network $(V,R_i)$, where
we will refer to it as the underlying graph of the association
scheme ($V$ is the vertex set of the association scheme which is
considered as vertex set of the underlying graph). In fact, an
association scheme partitions the relationships between pairs of
vertices into classes, so that for an arbitrary chosen vertex
(referred to as reference vertex), one can stratify the vertices
into distinct classes or stratas according to its relationships
with all of the other vertices. Moreover, this stratification is
independent of the choice of reference vertex. In the problem of
transfer of an arbitrary qubit state which is considered in this
work, we are given with $N$ spin-1/2 particles as the
corresponding vertex set; Then, for a given particle associated
with a vertex of the underlying graph, the strength of its
interaction with other $N-1$ particles is determined according to
its relationship with the other vertices defined via the relations
of the corresponding association scheme; In Refs.\cite{jss1},
\cite{jss2}, algebraic properties of association schemes have been
employed in order to evaluate the effective resistances in finite
resistor networks, where the relations of the corresponding
schemes define the kinds of resistances or conductances between
any two nodes of the networks. In Ref.\cite{dyn.}, a dynamical
system with $d$ different couplings has been investigated in which
the relationships between the dynamical elements (couplings) are
given by the relations between the vertexes according to the
corresponding association schemes. Indeed, according to the
relations $R_i$, the so-called adjacency matrices $A_i$ are
defined which form a commutative algebra known as Bose-Mesner (BM)
algebra. One of the important preferences of association schemes
is their useful algebraic structures that enables one to find the
spectrum of the adjacency matrices relatively easy; Then, for
different physical purposes, one can define particular spin
Hamiltonians which can be written in terms of the adjacency
matrices of an association scheme so that the corresponding
spectrums can be determined easily. Group association schemes are
particular schemes in which the vertices belong to a finite group
and the relations are defined based on the conjugacy classes of
the corresponding group. Working with these schemes is relatively
easy, since almost all of the needed information about the scheme,
for instance the so-called eigenvalue matrices $P$ and $Q$
associated with the scheme, can be obtained via the character
tables of the corresponding groups. We will use the technique of
the stratification of the underlying networks of group association
schemes (this technique can be used even for some graphs which are
not the underlying graphs of association schemes \cite{obata,obh},
not only for the purpose of state transfer, but also in
investigating the continuous time quantum walk (CTQW) over the
undirected graphs \cite{js}-\cite{krylov}) and employ their
algebraic structures in order to calculate the transition
probability amplitude between an arbitrary chosen reference vertex
(the identity element of the corresponding group is associated
with this vertex) and the antipode vertex (any element of the
center of the group which forms one element conjugacy class or one
element strata, can be considered as the corresponding antipode
vertex) and optimize it in order to attain PST between them
(transmission fidelity attains to maximum value one). As we will
see, the preference of this employment is that we are able to give
analytical formulas for coupling strengths in particular
Hamiltonians imposed to these graphs, in terms of the irreducible
characters of the corresponding group, in order for PST to be
achieved. In fact, we show that for those such networks in which
the group $G$ has non-trivial center (groups which have
non-trivial elements commuting with all of the other group
elements), an initial state encoded in one vertex of the network
(referred to as reference vertex or reference site) can be
transferred perfectly to the site labeled by any element of the
center of the corresponding group. We give explicit analytical
formula for the coupling strengths in terms of the parameters of
the corresponding group association scheme such as the diameter
$d$ of the scheme and the so-called first and second eigenvalue
matrices $P$ and $Q$ (which are given in terms of the irreducible
characters of the group). By calculating the optimal transmission
fidelity $F_{opt.}$, it is shown that for all such networks the
perfect transfer ($F_{opt.}=1$) can be achieved. We illustrate the
method for perfect transfer of a qubit over the considered
networks with details, and then generalize it to the PST of a
qudit over the same networks.

It should be noticed that, the interesting point of this work is
that, for the underlying graphs of group schemes considered in
this paper, the vertices of a the graphs are the elements of
particular finite groups, where the interactions between qubits
associated with vertices are governed by the relationship between
the vertices defined by the adjacency matrices $A_i$ of the
scheme; Therefore, according to different hamiltonians (different
kind of association schemes and consequently different kind of
relations or interactions) imposed to a given vertex set, one can
transfer a given state from a chosen vertex to different vertices.
For instance, consider a system of $8$ qubits each of which has
located at a corner of a hypercube; By imposing the interactions
between these $8$ vertices according to the relations of the group
scheme over $G=Z_2\times Z_2\times Z_2$, and taking some coupling
strengths between vertices equal to each other, one can obtain a
sub scheme which is the same as the Hamming scheme $H(2,3)$, and
transfer state of an arbitrary qubit initially prepared at the
vertex labeled by $|1\rangle\equiv|000\rangle$ to the antipode
vertex labeled by $|8\rangle\equiv|111\rangle$, perfectly;
However, by imposing the other relations defined by the other
group scheme with $8$ vertices, for instance the scheme over the
dihedral group $D_8$, one can transfer the same state initialized
at the vertex labeled by $|e\rangle\equiv|000\rangle$ to the
vertex labeled by $|a^2\rangle\equiv|011\rangle$ in the same
graph. In other words, for a given finite set of vertices, one can
associate different relationships between the vertices by choosing
different group association schemes in order to transfer a given
state from a chosen vertex to different vertices.

The organization of the paper is as follows: In section 2, we
recall some materials about association schemes, particularly the
group association schemes, their underlying networks and the
corresponding stratifications. Section $3$ is devoted to perfect
transfer of a qubit over the corresponding networks, where an
analytical formula for suitable set of coupling constants in
particular spin Hamiltonians is given. In section $4$, we
generalize the method to the perfect transfer of a qudit over the
same networks. Section $5$ contains two examples of the underlying
networks of group association schemes for which the PST is
achieved. The paper is ended with a brief conclusion.
\section{Underlying networks of association schemes} In this section, we review some
preliminary tools about the particular networks which are
considered through the paper. For material not covered in this
section, as well as more detailed information about association
schemes and their underlying networks, refer to \cite{Ass.sch.}
and \cite{js}.
\\
\textbf{Definition 1} Assume that $V$ and $E$ are vertex and edge
sets of a regular resistor network, respectively (each edge has a
certain conductance). Then, the relations $\{R_i\}_{0\leq i\leq
d}$
on $V\times V$ satisfying the following conditions\\
$(1)\;\ \{R_i\}_{0\leq i\leq d}$ is a partition of $V\times V$\\
$(2)\;\ R_0=\{(\alpha, \alpha) : \alpha\in V \}$\\
$(3)\;\ R_i=R_i^t$ for $0\leq i\leq d$, where
$R_i^t=\{(\beta,\alpha) :(\alpha, \beta)\in R_i\} $\\
$(4)$ For $(\alpha, \beta)\in R_k$, the number  $p^k_{i,j}=\mid
\{\gamma\in V : (\alpha, \gamma)\in R_i \;\ and \;\
(\gamma,\beta)\in R_j\}\mid$ does not depend on $(\alpha, \beta)$
but only on $i,j$ and $k$,\\ define a symmetric association scheme
of class $d$ on $V$ which is denoted by $Y=(V,\{R_i\}_{0\leq i\leq
d})$. Furthermore, if we have $p^k_{ij}=p^k_{ji}$ for all
$i,j,k=0,2,...,d$, then $Y$ is called commutative.

For examples of association schemes, consider a cube known as
Hamming scheme $H(3,2)$, in which $V$ (the vertex set) is the set
of $3$-tuples with entries in $F_2=\{0,1\}$. Two vertices are
connected if and only if they differ by exactly one entry (see
Fig. 1(a)). The distance between vertices, i.e. the length of the
shortest edge path connecting them, will then indicate which
relation they are contained in. E.g., if $x=(0,0,1)$, $y=(0,1,1)$
and $z=(1,0,1)$, then $(x,y)\in R_1$, $(x,z)\in R_1$ and $(y,z)\in
R_2$.

Let $Y=(V,\{R_i\}_{0\leq i\leq d})$ be a commutative symmetric
association scheme of class $d$, then the matrices
$A_0,A_1,...,A_d$ defined by
\begin{equation}\label{adj.}
    \bigl(A_{i})_{\alpha, \beta}\;=\left\{\begin{array}{c}
      \hspace{-2.3cm}1 \quad \mathrm{if} \;(\alpha,
    \beta)\in R_i, \\
      0 \quad \mathrm{otherwise} \quad \quad \quad(\alpha, \beta
    \in V) \\
    \end{array}\right.
\end{equation}
are adjacency matrices of $Y$ such that
\begin{equation}\label{ss}
A_iA_j=\sum_{k=0}^{d}p_{ij}^kA_{k}.
\end{equation}
From (\ref{ss}), it is seen that the adjacency matrices $A_0, A_1,
..., A_d$ form a basis for a commutative algebra \textsf{A} known
as the Bose-Mesner algebra of $Y$. This algebra has a second basis
$E_0,..., E_d$ (known as primitive idempotents of $Y$) so that
\begin{equation}\label{idem}
E_0 = \frac{1}{N}J, \;\;\;\;\;\;\ E_iE_j=\delta_{ij}E_i,
\;\;\;\;\;\;\ \sum_{i=0}^d E_i=I.
\end{equation}
where, $N:=|V|$ is the number of vertices (sites) and $J$ is the
all-one matrix in $\textsf{A}$. Let $P$ and $Q$ be the matrices
relating the two bases for $\textsf{A}$:
$$
A_i=\sum_{j=0}^d P_{ji}E_j, \;\;\;\;\ 0\leq j\leq d,
$$
\begin{equation}\label{m2}
E_i=\frac{1}{N}\sum_{j=0}^d Q_{ji}A_j, \;\;\;\;\ 0\leq j\leq d.
\end{equation}
Then clearly
\begin{equation}\label{pq}
PQ=QP=NI.
\end{equation}
It also follows that
\begin{equation}\label{eign}
A_iE_j=P_{ji}E_j,
\end{equation}
which shows that the $P_{ji}$ (resp. $Q_{ji}$) is the $j$-th
eigenvalue (resp. the $j$-th dual eigenvalue ) of $A_i$ (resp.
$E_i$) and that the columns of $E_j$ are the corresponding
eigenvectors. Thus, $m_i=$rank$(E_i)$ is the multiplicity of the
eigenvalue $P_{ji}$ of $A_i$ (provided that $P_{ji}\neq P_{jk}$
for $k \neq i$). We see that $m_0=1, \sum_i m_i=N$, and
$m_i=$trace$E_i=N(E_i)_{jj}$ (indeed, $E_i$ has only eigenvalues
$0$ and $1$, so rank($E_k$) equals the sum of the eigenvalues).

Clearly, each non-diagonal (symmetric) relation $R_i$ of an
association scheme $Y=(V,\{R_i\}_{_{0\leq i\leq d}})$ can be
thought of as the network $(V,R_i)$ on $V$, where we will call it
the underlying network of association scheme $Y$. In other words,
the underlying network $\Gamma=(V,R_1)$ of an association scheme
is an undirected connected network, where the set $V$ and $R_1$
consist of its vertices and edges, respectively. Obviously
replacing $R_1$ with one of the other relations such as $R_i$, for
$i\neq 0,1$ will also give us an underlying network
$\Gamma=(V,R_i)$ (not necessarily a connected network) with the
same set of vertices but a new set of edges $R_i$.
\subsection{Stratification} For an underlying network $\Gamma$, let
$W={\mathcal{C}}^n$ (with $n=|V|$) be the vector space over
$\mathcal{C}$ consisting of column vectors whose coordinates are
indexed by vertex set $V$ of $\Gamma$, and whose entries are in
$\mathcal{C}$. For all $\beta\in V$, let $\ket{\beta}$ denotes the
element of $W$ with a $1$ in the $\beta$ coordinate and $0$ in all
other coordinates. We observe that $\{\ket{\beta} | \beta\in V\}$
is an orthonormal basis for $W$, but in this basis, $W$ is
reducible and can be reduced to irreducible
$\textsf{A}$-submodules $W_i$, $i=0,1,...,d$ of the Bose-Mesner
algebra $\textsf{A}$ (by a $\textsf{A}$-submodule we mean a
subspace $W_i$ of $W$ such that $\textsf{A}W_i\subseteq W_i$),
i.e.,
\begin{equation}
W=W_0\oplus W_1\oplus...\oplus W_d,
\end{equation}
where, $d$ is diameter of the corresponding association scheme
(for more details see \cite{js}). If we define
 $\Gamma_i(o)=\{\beta\in V:
(o, \beta)\in R_i\}$ for an arbitrary chosen vertex $o\in V$
(called reference vertex), then, the vertex set $V$ can be written
as disjoint union of $\Gamma_i(\alpha)$, i.e.,
 \begin{equation}\label{asso1}
 V=\bigcup_{i=0}^{d}\Gamma_{i}(\alpha).
 \end{equation}
In fact, the relation (\ref{asso1}) stratifies the network into a
disjoint union of strata (associate classes) $\Gamma_{i}(o)$. With
each stratum $\Gamma_{i}(o)$ one can associate a unit vector
$\ket{\phi_{i}}$ in $W$ (called unit vector of $i$-th stratum)
defined by
\begin{equation}\label{unitv}
\ket{\phi_{i}}=\frac{1}{\sqrt{\kappa_{i}}}\sum_{\alpha\in
\Gamma_{i}(o)}\ket{\alpha},
\end{equation}
where, $\ket{\alpha}$ denotes the eigenket of $\alpha$-th vertex
at the associate class $\Gamma_{i}(o)$ and
$\kappa_i=|\Gamma_{i}(o)|$ is called the $i$-th valency of the
network ($\kappa_i:=p^0_{ii}=|\{\gamma:(o,\gamma)\in
R_i\}|=|\Gamma_{i}(o)|$). For $0\leq i\leq d$, the unit vectors
$\ket{\phi_{i}}$ of Eq.(\ref{unitv}) form a basis for irreducible
submodule of $W$ with maximal dimension denoted by $W_0$. Since
$\{\ket{\phi_{i}}\}_{i=0}^d$ becomes a complete orthonormal basis
of $W_0$, we often write
\begin{equation}
W_0=\sum_{i=0}^d\oplus \textbf{C}\ket{\phi_{i}}.
\end{equation}

Let $A_i$ be the adjacency matrix of the underlying network
$\Gamma$. From the action of $A_i$ on reference state
$\ket{\phi_0}$ ($\ket{\phi_0}=\ket{o}$, with $o\in V$ as reference
vertex), we have
\begin{equation}\label{Foc1}
A_i\ket{\phi_0}=\sum_{\beta\in \Gamma_{i}(o)}\ket{\beta}.
\end{equation}
 Then by using (\ref{unitv}) and (\ref{Foc1}),
 we obtain
\begin{equation}\label{Foc2}
A_i\ket{\phi_0}=\sqrt{\kappa_i}\ket{\phi_i}.
\end{equation}
\subsubsection{Group association schemes} Group
association schemes are particular association schemes for which
the vertex set contains elements of a finite group $G$ and the
relations $R_i$ are defined by \begin{equation}\label{relation}R_i
=\{(x, y)|yx^{-1}\in C_i\}, \;\ i=0,1,\ldots, d,\end{equation}
where $C_0 = \{e\},C_1, ...,C_d$ are the conjugacy classes of $G$.
Then, $(G,\{R_i\}_{0\leq i\leq d})$ becomes a commutative
association scheme, and it is called the group association scheme
of the finite group $G$. The $i$-th adjacency matrix $A_i$ is
defined as:
\begin{equation}\label{relation1}A_i={\bar{C_i}}:=\sum_{g\in C_i}g,\end{equation}
where $g$ is considered in the regular representation of the
group. Then, we can write
\begin{equation}{\bar{C_i}}
{\bar{C_j}}=\sum^{d}_{k=0}p^{k}_{ij}\bar{C_{k}}\end{equation} so
that the intersection numbers $p^{k}_{i,j}$, $i, j, k = 0, 1, ...,
d$ are given by \cite{7}
\begin{equation}p^{k}_{ij}=\frac{|Ci||Cj|}{|G|}\sum_{\chi}
\frac{\chi(\alpha_{i})\chi(\alpha_{j})\overline{\chi(\alpha_{k})}}{\chi(1)}\end{equation}
where the sum is taken over all the irreducible characters of $G$.
Therefore, the idempotents $E_0, ...,E_d$ of the group association
scheme are the projection operators as
\begin{equation}E_{k}=\frac{\chi_{k}(1)}{|G|}\sum_{\alpha\in G}\chi_{k}(\alpha^{-1})\alpha\end{equation}
Thus eigenvalues of adjacency matrices $A_{k}$ and idempotents
$E_{k}$ are
\begin{equation}\label{eqx}P_{ik}=\frac{\kappa_k\chi_{i}(\alpha_{k})}{d_{i}},\;\;\;\
Q_{ik}=d_{k}\overline{\chi_{k}(\alpha_{i})}\end{equation}
respectively, where $d_{j}=\chi_{j}(1)$ is the dimension of the
irreducible character $\chi_j$ and $\kappa_k\equiv|C_k|$ is the
$k$th valency of the graph.

It should be noticed that, in the cases that some of the conjugacy
classes are not real and hence some of the irreducible
representations are complex, we encounter with directed underlying
graphs or non-symmetric association schemes. In these cases, one
can form a symmetric association scheme out of a given
non-symmetric association scheme (see the appendix $A$ of
\cite{js}) so that, the transition probability amplitude between
the vertices $\phi_0\in C_0$ and $\beta\in C_k$ at time $t$ is
given by
\begin{equation}
\langle \beta|\phi_{0}(t)\rangle\;=\;\cases{ \frac{1}{|G|}\sum_i
d_{i}
e^{\frac{-i\kappa_1\chi_i(\alpha_1)t}{d_i}}\overline{\chi}_{i}(\beta)
& \mbox{for} \mbox{real} \mbox{representations} \cr
\frac{1}{|G|}\sum_i d_i
e^{\frac{-i\kappa_1(\chi_i(\alpha_1)+\overline{\chi}_i(\alpha_1))t}{d_i}}(\chi_i(\beta)+\overline{\chi}_{i}(\beta))
& \mbox{for} \mbox{complex} \mbox{representations}. \cr}
\end{equation}
for more details, see \cite{js}.

In the next section, we investigate state transfer over undirected
underlying graphs of group association schemes, where we will deal
with particular hamiltonians which are defined in terms of the
adjacency matrices of the corresponding association scheme; In the
other words, we will consider the underlying graphs of group
association schemes as spin networks in which the interactions
betweens spin states associated with the vertices of the graphs
are determined by the relations $R_i$ of the association scheme
defined as in Eq. (\ref{relation}) (or equivalently by the
adjacency matrices $A_i$ defined as in (\ref{relation1})).

As regards the above arguments, in the underlying graphs of
association schemes, the interactions between qubits associated
with vertices are governed by the relationship between the
vertices defined by the adjacency matrices $A_i$ of the scheme;
Therefore, according to different hamiltonians (different kind of
association schemes and consequently different kind of relations
or interactions) imposed to a given vertex set, one can transfer a
given state from a chosen vertex to different vertices. For
instance, consider a system of $8$ qubits each of which has
located at a corner of a hypercube; By imposing the interactions
between these $8$ vertices according to the relations of the group
scheme over $G=Z_2\times Z_2\times Z_2$, and taking some coupling
strengths between vertices equal to each other, one can obtain a
sub scheme which is the same as the Hamming scheme $H(2,3)$, and
transfer state of an arbitrary qubit initially prepared at the
vertex labeled by $\ket{1}\equiv\ket{000}$ to the antipode vertex
labeled by $\ket{8}\equiv\ket{111}$, perfectly (see Fig. 1(a));
However, as we will see in section $5$, by imposing the relations
defined by the group scheme over the dihedral group $D_8$, even by
considering only two types of relations (the coupling strengths
associated with other relations are taken to be zero) (see example
5.1 for details), we can transfer the same state initialized at
the vertex labeled by $\ket{e}\equiv\ket{000}$ to the vertex
labeled by $\ket{a^2}\equiv\ket{011}$ in the same graph (see Fig.1
(b)). In other words, for a given finite set of vertices, one can
associated different relationships between the vertices by
choosing different group association schemes in order to transfer
a given state from a chosen vertex to different vertices.
\section{Perfect state transfer of a qubit over antipodes of underlying networks of association schemes}
In order to set the scene, let us first recall the algorithm of
quantum state transfer over a general quantum network, briefly.
For more details see Refs. \cite{8,9'',PST,psd}.

A general finite quantum network is defined to be a simple,
connected, finite graph $\Gamma=(V,E)$, where $V$ denotes the
finite set of its vertices and $E$ the set of its edges. A
$2$-dimensional quantum system associated with such a graph is
defined by attaching a $2$-level (spin-$1/2$) particle to each
vertex of the graph so that with each vertex $i\in V$ one can
associate a Hilbert space ${\mathcal{H}}_i\simeq {\mathcal{C}}^2$.
The Hilbert space associated with $\Gamma$ is then given by
\begin{equation}
{\mathcal{H}}= \otimes_{_{i\in V}}{\mathcal{H}}_i =
({\mathcal{C}}^2)^{\otimes N},
\end{equation}
where $N:=|V|$ denotes the total number of vertices (sites) in
$\Gamma$. Now, suppose that we impose a particular spin
Hamiltonian $H$ (which governs to the interactions between the
spin-$1/2$ particles) to the graph so that the total $z$ component
of the spin, given by $\sigma^z_{tot}=\sum_{i\in V}\sigma^z_i$ is
conserved, i.e., $[\sigma^z_{tot},H]=0$. Hence the Hilbert space
${\mathcal{H}}$ decomposes into invariant subspaces, each of which
is a distinct eigenspace of the operator $\sigma^z_{tot}$. Then,
for the purpose of quantum state transfer, it suffices to restrict
our attention to the $N$-dimensional eigenspace of
$\sigma^z_{tot}$ corresponding to a spin configuration in which
all the spins except one are up and one spin is down. A basis
state for this eigenspace can hence be denoted by the $\ket{j}$,
where $j$ is the vertex in $\Gamma$ at which the spin is down.
Thus, $\{\ket{j}\; |\; j\in V\}$ denotes a complete set of
orthonormal basis vectors spanning the single down subspace.

The process of transmitting a quantum state from site $A$ to site
$B$ proceeds in two steps: initialization and evolution. First, a
quantum state $\ket{\psi}_A=\alpha\ket{0}_A+\beta\ket{1}_A\in
{\mathcal{H}}_A$ (with $\alpha,\beta\in \mathcal{C}$ and
$|\alpha|^2+|\beta|^2=1$) to be transmitted is created. The state
of the entire spin system after this step is given by
\be\label{eq1}\ket{\psi(t=0)}=\ket{\psi_A0...00_B}=\alpha\ket{0_A0...00_B}+\beta\ket{1_A0...00_B}=\alpha\ket{\b{0}}+\beta\ket{A},\ee
where, $\ket{\b{0}}:=\ket{0_A0...00_B}$, corresponds to the
configuration of all spins up (this state is a zero-energy
eigenstate of the considered Hamiltonian $H$). Then, the network
couplings are switched on and the whole system is allowed to
evolve under $U(t)=e^{-iHt}$ for a fixed time interval, say $t_0$.
The final state becomes  \be \ket{\psi(t_0)} =
\alpha\ket{\b{0}}+\beta\sum_{j=1}^Nf_{jA}(t_0)\ket{j} \ee where,
$f_{jA}(t_0):=\langle j|e^{-iHt_0}|A\rangle$. Any site $B$ is in a
mixed state if $|f_{AB}(t_0)|<1$, which also implies that the
state transfer from site $A$ to $B$ is imperfect. In this paper,
we will focus only on PST. This means that we consider the
condition \be\label{eq3} |f_{AB}(t_0)|=1\;\;\ \mbox{for}\;\
\mbox{some}\;\ 0<t_0<\infty\ee
 which can be interpreted as the signature of
perfect communication (or PST) between $A$ and $B$ in time $t_0$.
The effect of the modulus in (\ref{eq3}) is that the state at $B$,
after transmission, will no longer be $\ket{\psi}$, but will be of
the form \be \alpha\ket{0}+e^{i\phi}\beta\ket{1}. \ee The phase
factor $e^{i\phi}$ is not a problem because $\phi$ is independent
of $\alpha$ and $\beta$ and will thus be a known quantity for the
graph, which we can correct for with an appropriate phase gate
(for more details see for example \cite{8,9'',yung,yung1}).

The model we will consider is an underlying network of a group
association scheme over a finite group $G$ consisting of $N=|G|$
sites labeled by $\{1,2, ... ,N\}$ and diameter $d$. Then we
stratify the network with respect to a chosen reference site, say
$1=e$ (the unit element of the group), and assume that the group
$G$ has a non trivial conjugacy class, say $C_m$, with cardinality
one which contains the output site $N$ (i.e.,
$\ket{\phi_m}=\ket{N}$). At time $t=0$, the qubit in the first
(input) site of the network is prepared in the state
$\ket{\psi_{in}}$. We wish to transfer the state to the $N$th
(output) site of the network with unit efficiency after a
well-defined period of time. The standard basis for an individual
qubit (e.g., one can consider state of a spin 1/2 particle) is
chosen to be $\{|0\rangle=|\downarrow\rangle,\;\
|1\rangle=|\uparrow\rangle\}$, and we shall assume that initially
all spins point ``down" along a prescribed $z$ axis; i.e., the
network is in the state $|\b{0}\rangle=|0_A00...00_B\rangle$.
Then, we consider the dynamics of the system to be governed by the
quantum-mechanical Hamiltonian
\begin{equation}
\label{H}H=\sum_{l=0}^dJ_l\sum_{(i,j)\in R_l}\vec{\sigma}^{(i)}\cdot \vec{\sigma}^{(j)}
\end{equation}
where, ${\mathbf{\sigma}}_i$ is a vector with familiar Pauli
matrices $\sigma^x_i, \sigma^y_i$ and $\sigma^z_i$  as its
components acting on the one-site Hilbert space
${\mathcal{H}}_i\simeq {\mathcal{C}}^2$ associated with the site
$i\in V$, and $J_{l}$ is the coupling strength between the
reference site $1$ and all of the sites belonging to the $l$-th
stratum with respect to $1$. As it can be easily seen, the
Hamiltonian (\ref{H}) commutes with the total spin operator
(conservation). That is, since we have $[\sigma^z_{tot},H]=0$,
hence the total Hilbert space ${\mathcal{H}}=\otimes_{_{i\in
V}}{\mathcal{H}}_i = ({\mathcal{C}}^2)^{\otimes N}$ decomposes
into invariant subspaces, each of which is a distinct eigenspace
of the operator $\sigma^z_{tot}$. Then, as it was shown in
Ref.\cite{PST}, one can rewrite the Hamiltonian (\ref{H}) in the
single down subspace with basis vectors
$\ket{i}=\ket{\uparrow...\uparrow\underbrace{\downarrow}_i\uparrow...\uparrow}$,
$i=1,2,\ldots, N$ as follows \be\label{HA'}H=2\sum_{i=0}^d
J_iA_i+\frac{N-4}{2}\sum_{i=0}^d\kappa_iJ_i\mathbf{I}.\ee Then, by
using (\ref{m2}), we have
$$e^{-iHt}=e^{-it\frac{N-4}{2}\sum_{i=0}^d\kappa_iJ_i}\sum_{k=0}^de^{-2it\sum_{i=0}^dJ_iP_{ki}}E_k.$$
Now, let the $m$-th stratum of the network contains only one
vertex, then one can write
$$\langle\phi_m|e^{-iHt}|\phi_0\rangle=\sum_{k=0}^de^{-it\sum_{i=0}^dJ_iP_{ki}}\langle\phi_m|E_k|\phi_0\rangle=$$
\begin{equation}\label{amp.}\frac{1}{|G|}\sum_{k=0}^de^{-it\sum_{i=0}^dJ_iP_{ki}}Q_{mk}=\frac{1}{|G|}\sum_{k=0}^de^{-it\sum_{i=0}^d\frac{J_i\kappa_i}{d_k}\chi_k(\alpha_i)}d_k\bar{{\chi}}_k(\alpha_m).
\end{equation}
In order that the optimal state transfer be achieved, the above
probability amplitude must take its maximum value. By using
(\ref{amp.}), we obtain
$$\max|\langle\phi_m|e^{-iHt}|\phi_0\rangle|^2=\frac{1}{|G|^2}\max |\sum_k|Q_{mk}|e^{-it\sum_iJ_iP_{ki}+i\varepsilon_k\pi}|^2$$
where, $\epsilon_k$ is $0$ or $1$ depending on the sign of
$Q_{mk}$ (the sign of $\chi_k(\alpha_m)$). The above maximum
attains if we have
\be\label{1}-t\sum_iJ_iP_{ki}+\varepsilon_k\pi=\phi+2l\pi\;\;\
\rightarrow \;\ (PJ)_k=\frac{\varepsilon_k\pi-\phi-2l\pi}{t_0},\;\
l\in {\mathrm{Z}} \ee ($\phi$ is a constant phase). Then, by using
(\ref{pq}), (\ref{eqx}) and (\ref{1}), the optimal coupling
constants (for which the optimal state transfer is achieved) are
obtained as
\begin{equation}\label{coupl.}J_k=\frac{1}{|G|}\sum_{i=0}^d\frac{\varepsilon_i\pi-\phi-2l\pi}{t_0}Q_{ki}=\frac{1}{|G|}\sum_{i=0}^d\frac{\varepsilon_i\pi-\phi-2l\pi}{t_0}d_i\bar{\chi}_i(\alpha_k).\end{equation}
so that the optimal fidelity is given by
\be\label{finres}F_{opt.}=\max\langle\phi_m|e^{-iHt}|\phi_0\rangle=\frac{1}{|G|}\sum_{i=0}^d
|Q_{mi}|=\frac{1}{|G|}\sum_{i=0}^dd_i|\bar{\chi}_i(\alpha_m)|.\ee
From the fact that $\alpha_m\in C_m$ belongs to the center of the
group (and so commute with all elements of the group), the Schur's
lemma implies that the irreducible representation of $\alpha_m$ is
equal to $\mu \mathbf{1}$ for some $\mu\in \mathcal{C}$. Now, let
$a$ be the order of $\alpha_m\in G$ ($\alpha^a_m=1$). Then,
clearly $\mu $ must be the $a$th root of unity, i.e., we have
$\mu=e^{2\pi i/a}$ and so $|\mu|=1$. Then, we will have
$|\chi_k(\alpha_m)|=|\mu|d_k=d_k$ and the optimal fidelity attains
\be\label{finres1}F_{opt.}=\frac{1}{|G|}\sum_{k=0}^dd^2_k=1.\ee
where we have used the well known identity $\sum_{k=0}^dd^2_k=|G|$
from group theory. The above result indicates that all finite
groups with non-trivial center can allow state transfer with
optimal fidelity one, i.e., perfect state transfer.
\section{Generalization to perfect state transfer of a $D$-level quantum state}
In the $D^N$ dimensional Hilbert space associated with system of
$N$, $D$-level quantum states (${\mathcal{H}}\simeq
({\mathcal{C}}^D)^{\otimes N}$), the Hamiltonian (\ref{H}) can be
generalized as
\begin{equation}\label{Hd}H=\sum_{l=0}^{d}J_l\sum_{(i,j)\in
R_l}\vec{\lambda}_{i}\cdot \vec{\lambda}_{j},\end{equation} where,
${\mathbf{\vec{\lambda}}}_i$ is a $D^2-1$ dimensional vector with
generators of $SU(D)$ as its components acting on the one-site
Hilbert space ${\mathcal{H}}_i\simeq {\mathcal{C}}^{D}$.

Let us denote a state in which the $i$-th site has been exited to
the level $\nu$ by
$\ket{\nu_i}\equiv\ket{0\ldots\ldots0\underbrace{\nu}_i0\ldots0}$.
Then, the hamiltonian $H$ can be diagonalized in each subspace
$S^{(\nu)}$ spanned by the vectors $\ket{\nu_i}$, $i=1,\ldots, N$,
for $\nu=1,...,D-1$ (for more details see \cite{psd}). If we call
the states with only one site excited, as one particle states and
the subspace spanned by these vectors comprise the one particle
sector of the full Hilbert space, then, the whole one particle
subspace $\mathcal{S}$ can be written as
$${\mathcal{S}}={\mathcal{S}}^{(1)}\oplus {\mathcal{S}}^{(2)}\oplus\ldots\oplus {\mathcal{S}}^{(D-1)}.$$
In the other words, in $D^N$ dimensional Hilbert space
$\mathcal{H}$, we deal with $D-1$ one particle subspaces (recall
that, each of these subspaces has dimension $N$). In the case of a
system of $N$ qubit ($D=2$), we have only one one-particle
subspace of dimension $N$.

As it was shown in Refs. \cite{PST}, \cite{psd}, the hamiltonian
(\ref{Hd}) restricted to each one particle subspace $S^{(\nu)}$,
for $\nu=1,2,\ldots,d-1$, can be written in terms of the adjacency
matrices $A_l$ as
\begin{equation}\label{HAd}H=2\sum_{l=0}^{d}J_lA_l+\frac{(N-2)D-4N}{2D}\sum_{l=0}^{d}J_l\kappa_l
\mathbf{1}.\end{equation} Then the same procedure can be applied
in order to obtain the optimal fidelity. The Eq.(\ref{HAd}) is
similar to the Eq. (\ref{HA'}) apart from the non-important second
term which is multiple of the identity. Therefore, maximizing the
probability amplitude $\langle \phi_m|e^{-iHt}|\phi_0\rangle$
gives the same constraints on coupling constants $J_l$ and leads
to the same optimal fidelity (\ref{finres1}).
\section{Examples} In this section,
we consider two examples of group association schemes and
investigate PST over them.
\subsection{The dihedral group $D_{2n}$}
The dihedral group $G = D_{2n }$ is generated by two generators
$a$ and $b$ with the following relations:
\begin{equation}D_{2n}=\langle a,b:a^{n}=1,b^{2}=1,b^{-1}ab=a^{-1}\rangle\end{equation}
We consider the case of even $n=2m$, the case of odd $n$ can be
considered similarly. The Dihedral group $D_{2n}$ with even $n=2m$
has $m+3$ conjugacy class so that $C_m=\{a^m\}$. For even $n=2m$,
the $m+3$ conjugacy classes are given by
$$C_0=\{1\},\;\  C_r=\{a^r,a^{-r}\}\;\ ; \;\ 1\leq r \leq m-1 , \;\ C_m=\{a^m\},\;\ C_{m+1}=\{a^{2j}b\;\ ; \;\ j=0,\ldots, m-1\},$$
$$ C_{m+2}=\{a^{2j+1}b\;\ ; \;\ j=0,\ldots, m-1\}.$$
The character table is given by
$$\begin{tabular}{|c|c|c|c|c|c|}
  \hline
   $D_{2n}$& $e$ & $a^{m}$ & $a^{r}\;\ (1\leq r\leq m-1)$ & $b$ & $ab$ \\
  \hline
  $\chi_{0}$ & 1 & 1 & 1 & 1 & 1 \\
  $\chi_{1}$ & 1 & 1 & 1 & -1 & -1 \\
  $\chi_{2}$ & 1 & $(-1)^{m}$ & $(-1)^{r}$ & 1& -1 \\
  $\chi_{3}$ & 1 & $(-1)^{m}$ & $(-1)^{r}$ & -1 & 1 \\
  $\psi_{j} \;\ (1\leq j\leq m-1)$ & 2 & $2(-1)^j$ & $2\cos(2\pi jr/n)$ &  0 & 0 \\
  \hline
\end{tabular}$$
Then, by using the result (\ref{finres}) we obtain
$$F_{opt.}= \frac{1}{|G|}\sum_{k=0}^{m+2}d_k|\bar{\chi}_k(\alpha_m)|=\frac{1}{4m}\{4+4(m-1)\}=1.$$
From (\ref{coupl.}), one can evaluate the optimal coupling
constants $J_l$. For example, for $n=4$ we obtain
$$J_0=J_2=J_4=0, \;\ J_1=\frac{\pi}{2t_0},\;\ J_3=\frac{2\pi}{t_0},$$
where the above values are obtained by choosing $\phi=\pi$,
$l_0=3$, $l_1=-1$, $l_2=-l_3=2$ and $l_4=0$ in the Eq.
(\ref{coupl.}). It is seen that by choosing $\phi=\pi$, all
coupling constants become zero except for $J_1$ and $J_3$ (see
Fig. 1(b)).
\subsection{The Cubic group $T_{h}$}
One of the point groups with high or polyhedral symmetry is the
cubic group $T_h$ of order $24$ with pyritohedral symmetry. This
group is isomorphic to $A_4 \times C_2$, where $A_4$ is the
alternating group of order $12$. It is the symmetry of a cube with
on each face a line segment dividing the face into two equal
rectangles, such that the line segments of adjacent faces do not
meet at the edge. The symmetries correspond to the even
permutations of the body diagonals and the same combined with
inversion. The group $T_h$ has $8$ conjugacy classes with the
following character table
$$\begin{tabular}{|c|c|c|c|c|c|c|c|c|}
  \hline
   $T_h $& $C_0$ & $C_1$ & $C_2$ & $C_3$ & $C_4$ & $C_5$ & $C_6$ & $C_7$ \\
  \hline
  $\chi_{0}$ & 1 & 1 & 1 & 1 & 1 &1 & 1&1 \\
  $\chi_{1}$ & 1 & 1 & -1 & -1 & 1 & 1&-1 &-1 \\
  $\chi_{2}$ & 1 & 1& 1 & 1 & $\omega$ & $\omega^2$  & $\omega$ & $\omega^2$ \\
  $\chi_{3}$ & 1 & 1 & 1 & 1 & $\omega^2$  & $\omega$ & $\omega^2$ & $\omega$ \\
  $\chi_{4}$ & 1 & -1 & -1 & -1 & $\omega$ & $\omega^2$  & $-\omega$ & $-\omega^2$ \\
  $\chi_{5}$ & 1 & 1 & -1& -1 & $\omega^2$  & $\omega$ & $-\omega^2$ & $-\omega$ \\
  $\chi_{6}$ & 3 & -1 & 3 & -1 & 0 & 0& 0& 0\\
  $\chi_{7}$ & 3 & -1 & -3 & 1 & 0 & 0& 0& 0\\
  \hline
\end{tabular}$$
with $\omega:=e^{2\pi i/3}$. By combining the classes $C_4$ and
$C_5$ and denoting it by $\tilde{C_4}= C_4\cup C_5$, and $C_6$
with $C_7$ and denoting the new obtained class as $\tilde{C_5}$,
one can obtain a symmetric association scheme with undirected
underlying graph. Now, using the result (\ref{coupl.}) one can
evaluate the optimal coupling constants $J_l$ for, $l=0,1,\ldots,
5$ as follows:
$$J_0=J_1=J_2=J_4=0,\;\ J_3=J_5=\frac{\pi}{2t_0},$$
where, we have chosen $\phi=\pi/2$, $l_0=-l_1=3$ and
$l_2=l_3=l_4=l_5=0$ in the Eq. (\ref{coupl.}). By these choices,
all coupling constants become zero except for $J_3$ and $J_5$ (see
Fig.2).
\subsection{The Clifford group}The Clifford algebra with $n$ generator matrices
$\gamma_1,\gamma_2,...,\gamma_n$, obeys the following relations
\cite{3}
\begin{equation}\gamma_i\gamma_j+\gamma_j\gamma_i=2\delta_{ij}I\end{equation}
Thus, the $\gamma$'s have square $1$ and anti-commute. The
Clifford group denoted by $CL(n)$ has $2^{n+1}$ elements as
$$CL(n)=\{\pm1, \pm\gamma_{i_{1}} . . .\gamma_{i_{j}}; \;\  i_ 1 < . . . < i_j , j=1,\ldots, n\},$$
where, $i_r\in \{1,2,\ldots, n\}$. We suppose $n > 2 $ throughout.
It is well known that \cite{3}, the center of $CL (n)$ denoted by
$ Z (CL( n))$, consists of $\{\pm1\}$ if $n$ is even and $ \{
\pm1,\pm\gamma_1 ...\gamma_n\} $ if $n$ is odd. $CL( n)$ has $
2^n$ one-dimensional representations, each real. In each such
representation, $U( -1) =I$; Any irreducible representation with
dimension greater than $1$ has $U(-1) =-I$.

For even $n$, the conjugacy classes are given by
$$C_0=\{1\},\;\ C_1=\{-1\},\;\ C_2=\{\gamma_{1},-\gamma_{1}\},\ldots,\;\ C_j=\{\gamma_{i_1}...\gamma_{i_j},-\gamma_{i_1}...\gamma_{i_j}\},$$
$$C_{2^n+1}=\{\gamma_{1}...\gamma_{n},-\gamma_{1}...\gamma_{1}\},$$
whereas for odd $n$, we have
$$C_0=\{1\},\;\ C_1=\{-1\},\;\ C_2=\{\gamma_{1},-\gamma_{1}\},\ldots,\;\ C_j=\{\gamma_{i_1}...\gamma_{i_j},-\gamma_{i_1}...\gamma_{i_j}\},$$$$C_{2^n+1}=\{\gamma_{1}...\gamma_{n}\},\;\ ,C_{2^n+2}=\{-\gamma_{1}...\gamma_{1}\}.$$
In the following we consider only the case of even $n$, the case
of odd $n$ can be considered similarly.

The characters of the $2^n$ one dimensional representations are
given by
$$\chi_k(1)=\chi_k(-1)=1,\;\;\ 0\leq k\leq 2^n-1,$$
$$\chi_{2^n}(\pm \gamma_A)=\pm \delta_{A\emptyset}2^{n/2}\;\ \Rightarrow  \;\ \chi_{2^n}(1)=2^{n/2},\;\ \chi_{2^n}(-1)=-2^{n/2}. $$
Then, by using the result (\ref{finres}) we obtain
$$F_{opt.}= \frac{1}{2^{n+1}}\sum_{k=0}^{2^n}d_k|\bar{\chi}_k(\alpha_m)|=\frac{1}{2^{n+1}}\{2^{n}+2^{n/2}(2^{n/2})\}=1.$$

Even though we considered only two examples of PST over underlying
networks of group association schemes in details, for all
underlying networks of group association schemes for which the
corresponding group has non-trivial center, PST can be achieved.
For example, underlying networks of group association schemes
associated with abelian finite groups and all $p$-groups (a
$p$-group is a group whose order is a power of the prime number
$p$; the centers corresponding to the $p$-groups are non-trivial
\cite{7}) allow PST. It should also be noticed that for direct
product groups $G=G_1\times G_2$, so that PST can be achieved for
$G_1$ and $G_2$ at the same time $t_0$, PST will be achievable for
the group $G$, since the transition amplitudes $\langle
\phi_k|e^{-iHt}|\phi_0\rangle$ corresponding to the underlying
network associated with group association scheme over $G$ are
product of the amplitudes corresponding to the groups $G_1$ and
$G_2$ (for more details see \cite{js}). Particularly, for any
finite group $G$ with non-trivial center, not only PST over
underlying network of group association scheme over $G$ is
achieved, but also it is achievable for direct product group
$\underbrace{G\times G\times \ldots \times G}_n$. For instance,
the $n$-dimensional hypercube network ($n$-cube) can be viewed as
the underlying network of group association scheme over the
product group $\underbrace{Z_2\times Z_2\times \ldots \times
Z_2}_n$ and PST over it can be achieved by only nearest neighbor
coupling strengths.
\section{Conclusion}
Perfect state transfer of a qudit over antipodes of the so called
underlying networks of group association schemes, was
investigated, where by using the group properties and the
algebraic structure of these networks (such as Bose-Mesner
algebra), an explicit analytical formula for coupling constants in
the Hamiltonians (in terms of the irreducible characters of the
corresponding group) was given, so that the state of a particular
qudit initially encoded on one site can be perfectly evolved to
the opposite site without any dynamical control.
\newpage
\setcounter{section}{0} \setcounter{equation}{0}
\renewcommand{\theequation}{A-\roman{equation}}
{\Large{Appendix A}}\\
As regards the arguments of section $3$, for the purpose of the
perfect state transfer, we consider the underlying graphs of group
association schemes in which the corresponding group has
non-trivial center in order to the group have no-trivial one
element conjugacy classes or stratas with size one, i.e., the last
stratum of the graph denoted by $\Gamma_m(o)$ contains only one
site ($\kappa_m=|\Gamma_m(o)|=1$). Then, by imposing the
constraint that the transition probability amplitude
$\langle\phi_m|e^{-iHt}|\phi_0\rangle$ be equal to an arbitrary
phase such as $e^{i\theta}$ and consequently the transition
probability amplitudes $\langle\phi_i|e^{-iHt}|\phi_0\rangle$ be
zero, for all $i\neq m$, one can achieve PST from the reference
site $\ket{\phi_0}=\ket{e}$ associated with the first stratum to
the site of the last stratum $\ket{\phi_m}$.

Even though, as it was shown in section 3, we need to evaluate
only the probability amplitude
$\langle\phi_m|e^{-iHt}|\phi_0\rangle$ in order for the
corresponding fidelity to be maximized and PST be attained, here
we give the transition probability amplitudes between the
reference site $\ket{e}$ and any group element $\beta\in G$, in
terms of the coupling constants and irreducible characters of the
group which can be used in some areas of quantum information such
as generating quantum entanglements between any two sites of a
given finite network. To do so, given a finite group $G$, let
$\beta \in G$ belong to the $l$-th conjugacy class of the group.
In general, assume that the group $G$ has $R$ real conjugacy
classes ($C_i=C^{-1}_i$, for $i=0,1,\ldots, R-1$) and so it will
posses $R$ real representations; Now, we use the stratification
and algebraic properties of the underlying graph of the group
association scheme over the group $G$, to write
\begin{equation}\label{amp.}\langle \beta|e^{-iHt}|\phi_0\rangle=\frac{1}{\sqrt{\kappa_l}}\langle \phi_l|e^{-iHt}|\phi_0\rangle
=\frac{1}{|G|\sqrt{\kappa_l}}\{\sum_{k=0}^{R-1}e^{-it\Theta_{\chi_k}}d_k
\bar{\chi_k}(\alpha_l)+\sum_{k=R}^{\frac{d+1-R}{2}}e^{-it\Theta_{\chi_k}}d_k
(\bar{\chi_k}(\alpha_l)+\chi_k(\alpha_l))\},\end{equation}where
\begin{equation}\label{teta}\Theta_{_{\chi_k}}:=\sum_{l=0}^{R-1}\frac{J_l\kappa_l}{d_k}\chi_k(\alpha_l)+\sum_{l=R}^{\frac{d+1-R}{2}}\frac{J_l\kappa_l}{d_k}(\chi_k(\alpha_l)+\bar{\chi_k}(\alpha_l))\end{equation}
where we have used the fact that
$\chi_k(\bar{\alpha_l})=\bar{\chi_k}(\alpha_l)$. As it is seen
from (\ref{teta}), we have
$\Theta_{_{\chi_k}}=\Theta_{\bar{\chi_k}}$.

It should be also noticed that, for groups $G$ with non-real
conjugacy classes, the corresponding group association schemes are
not symmetric and so the corresponding underlying graphs are not
undirected; Therefore, for the purpose of state transfer, we need
to symmetrize the group schemes in order to obtain undirected
graphs. Now, we recall that for any non real conjugacy class $C_i$
($C_i\neq C^{-1}_i$), $C^{-1}_i$ is also a conjugacy class of $G$
with the same size of $C_i$, i.e., $|C_i|=|C^{-1}_i|$; Then, as in
Ref. \cite{js} was shown, the symmetrized classes $C_i$, for
$i=0,\ldots, R-1$; $\tilde{C}_i\equiv C_i\cup C^{-1}_i$, for
$i=R,\ldots, \frac{d+1-R}{2}$, form a symmetric association scheme
which is a sub scheme of the non-symmetric group association
scheme over $G$. Now, by choosing the identity element of the
group as reference vertex and stratify the underlynig graph with
respect to it, the resulted symmetrized sub scheme can be
considered for the purpose of PST if its last stratum denoted by
$\Gamma_m(e)$ has size one, i.e., we must have $|C_m=C_m^{-1}|=1$
in order for PST to be studied.

\newpage
{\bf Figure Captions}

{\bf Figure 1:} (a) Denotes a vertex set $\{1,2,\ldots, 8\}$ with
interactions according to the relations of Hamming scheme
$H(3,2)$, where with only one non-zero coupling strengths $J_1$,
PST between the vertex $(000)$ and the vertex $(111)$ is achieved.
(b) Denotes the same vertex set $\{1,2,\ldots, 8\}$ with
interactions according to the relations of the group association
scheme over dihedral group $D_8$, where with only two non-zero
coupling strengths $J_1$ and $J_3$, PST from the vertex $e$ to the
vertex $a^2$ is achieved; The solid lines denote the interaction
coupling $J_1$ and the dashed lines denote the coupling strength
$J_3$.

{\bf Figure 2:} Shows the underling graph of the symmetrized
scheme obtained from the group scheme over the Cubic group $T_h$,
where non-zero coupling strengths $J_3=J_5$ and
$J_0=J_1=J_2=J_4=0$, PST from $C_0$ to $C_2$ is achieved.
\end{document}